\documentclass[reprint,showpacs,preprintnumbers,amsmath,amssymb,aps,pre]{revtex4-1}

\usepackage{graphicx}
\usepackage{dcolumn}
\usepackage{bm}
\usepackage[OT4]{fontenc} 

\newcommand{\rozmiar}{0.47}

\begin{document}

\title{On the phase diagram of the zero-bandwidth extended Hubbard model\\ with intersite magnetic interactions for strong on-site repulsion limit}
\author{Szymon Murawski}
\author{Konrad Kapcia}%
    \email{corresponding author; e-mail: \url{kakonrad@amu.edu.pl}}
\author{Grzegorz Paw\l{}owski}
\author{Stanis\l{}aw Robaszkiewicz}%
\affiliation{Electron States of Solids Division, Faculty of Physics, Adam Mickiewicz University, Umultowska 85, 61-614 Pozna\'n, Poland
}%

\date{May 27, 2011}

\begin{abstract}
In this report we have analyzed a~simple effective model for a~description of magnetically ordered insulators. The Hamiltonian considered consists of the effective on-site interaction ($U$) and the intersite Ising-like magnetic exchange interaction ($J$) between nearest neighbors. For the first time the phase diagrams of this model have been determined within Monte Carlo simulation on 2D-square lattice. They have been compared with results obtained within variational approach, which treats the on-site term exactly and the intersite interactions within mean-field approximation. We show within both approaches that, depending on the values of interaction parameters and the electron concentration, the system can exhibit not only homogeneous phases: \mbox{(anti-)ferromagnetic} (F) and nonordered (NO), but also phase separated states (PS: \mbox{F--NO}).
\end{abstract}

\pacs{71.10.Fd, 75.30.Fv, 64.75.Gh, 71.10.Hf}
\maketitle


\section{Introduction}

The extended Hubbard model with spin exchange interaction \cite{MRR1990,JM2000,DJSZ2004,CzR2006} is a conceptually simple effective model for a~description of magnetically ordered insulators in narrow band systems.

In this report we will focus on the zero-bandwidth limit of the extended Hubbard model with magnetic Ising-like interactions, which has the following form:
\begin{equation}\label{row:1}
\hat{H} = U\sum_i{\hat{n}_{i\uparrow}\hat{n}_{i\downarrow}} - 2J\sum_{\langle i,j\rangle}{\hat{s}^z_{i}\hat{s}^z_{j}} - \mu\sum_{i}{\hat{n}_{i}},
\end{equation}
where $U$~is the on-site density interaction,  $J$~is $z$-component of the intersite magnetic exchange interaction, \mbox{$\sum_{\left\langle i,j\right\rangle}$} restricts the summation to nearest neighbors. $\hat{c}^{+}_{i\sigma}$ denotes the creation operator of an electron with spin $\sigma$ at the site $i$, \mbox{$\hat{n}_{i}=\sum_{\sigma}{\hat{n}_{i\sigma}}$}, \mbox{$\hat{n}_{i\sigma}=\hat{c}^{+}_{i\sigma}\hat{c}_{i\sigma}$} and \mbox{$\hat{s}^z_i = \frac{1}{2}(\hat{n}_{i\uparrow}-\hat{n}_{i\downarrow})$}. The chemical potential $\mu$ depending on the concentration of electrons is calculated from
\begin{equation}\label{row:2}
n = \frac{1}{N}\sum_{i}{\left\langle \hat{n}_{i} \right\rangle},
\end{equation}
with \mbox{$0\leq n \leq2$} and $N$ is the total number of lattice sites.

The hamiltonian (\ref{row:1}) can be considered as a~very simplified model for the family of A$_{0.5}$M$_2$X$_4$ compounds (where A is Ga or Al, M is one of the transition metals V or Mo, and X is S, Se, or Te). These compounds exhibit very interesting ferromagnetic behavior which is a~mixture of itinerant and localized behavior~\cite{b1,b2}. However, the single-particle excitations play a dominant role in the magnetic behavior of these compounds. Although the electrons are not itinerant in the system, there is a finite density of states at the Fermi level, and therefore low energy charge excitations are possible~\cite{b3}. This points that the magnetic properties result from this band of localized electrons.

The model (\ref{row:1}) can be treated as an effective model of magnetically ordered insulators. The interactions $U$ and $J$ will be treated as effective ones and  be assumed  to include all the possible contributions and renormalizations like those coming from the strong electron-phonon coupling or from the coupling between electrons and other electronic subsystems in solid or chemical complexes. In such a general case arbitrary values and signs of $U$ are important to consider.
One should notice that ferromagnetic (\mbox{$J>0$}) interactions are simply mapped onto the antiferromagnetic cases (\mbox{$J<0$}) by redefining the spin direction on one sublattice in lattices decomposed into two interpenetrating sublattices. Thus, we restrict ourselves to the case \mbox{$J>0$}.

For the model (\ref{row:1}) only the ground state phase diagram as a function of $\mu$~\cite{BS1986} and special cases of half-filling (\mbox{$n=1$})~\cite{R1979} have been investigated till now. Some  our preliminary results have been also presented in Ref.~\onlinecite{KKR2010}.

We have performed extensive study of the phase diagrams of the model (\ref{row:1}) for arbitrary $n$ and $\mu$ \cite{KKR2010,WK2009,KKM0000}. In the analysis we have adopted two complementary methods: (i)~a~variational approach (VA), which treats the on-site interaction term ($U$) exactly and the intersite interactions ($J$) within the mean-field approximation (MFA) and (ii)~Monte Carlo (MC) simulations for $d=2$ dimensional square (SQ) lattice in the grand canonical ensemble. In this report we present some results concerning strong on-site repulsion limit.

The ferromagnetic (F) phase is  characterized by non-zero value of the magnetic order parameter (magnetization) defined as  $m=(1/N)\sum_{i}\langle\hat{s}^z_i\rangle$. In the nonordered (NO) phase \mbox{$m=0$}.

The phase separation (PS) is a state in which  two domains with different electron concentration exist
(coexistence of two homogeneous phases).  In the model considered only one PS state (\mbox{F--NO}) can occur.

In the paper we have used the following convention. A~second (first) order transition is a~transition between homogeneous phases  with a~(dis-)continuous change of the order parameter at the transition temperature. A~transition between homogeneous phase and PS state is symbolically named as a~``third order'' transition. During this transition a~size of one domain in the PS state decreases continuously to zero at the~transition temperature. One should notice that the first order transition line on the diagrams for fixed $\mu$ splits into  two ``third order'' lines and it is connected with occurrence of PS sates on the diagrams for fixed $n$.

We also introduce the following denotation: \mbox{$J_0=zJ$}, where $z$ is the number of nearest neighbors.

\begin{figure}
    \centering
    \includegraphics[width=\rozmiar\textwidth]{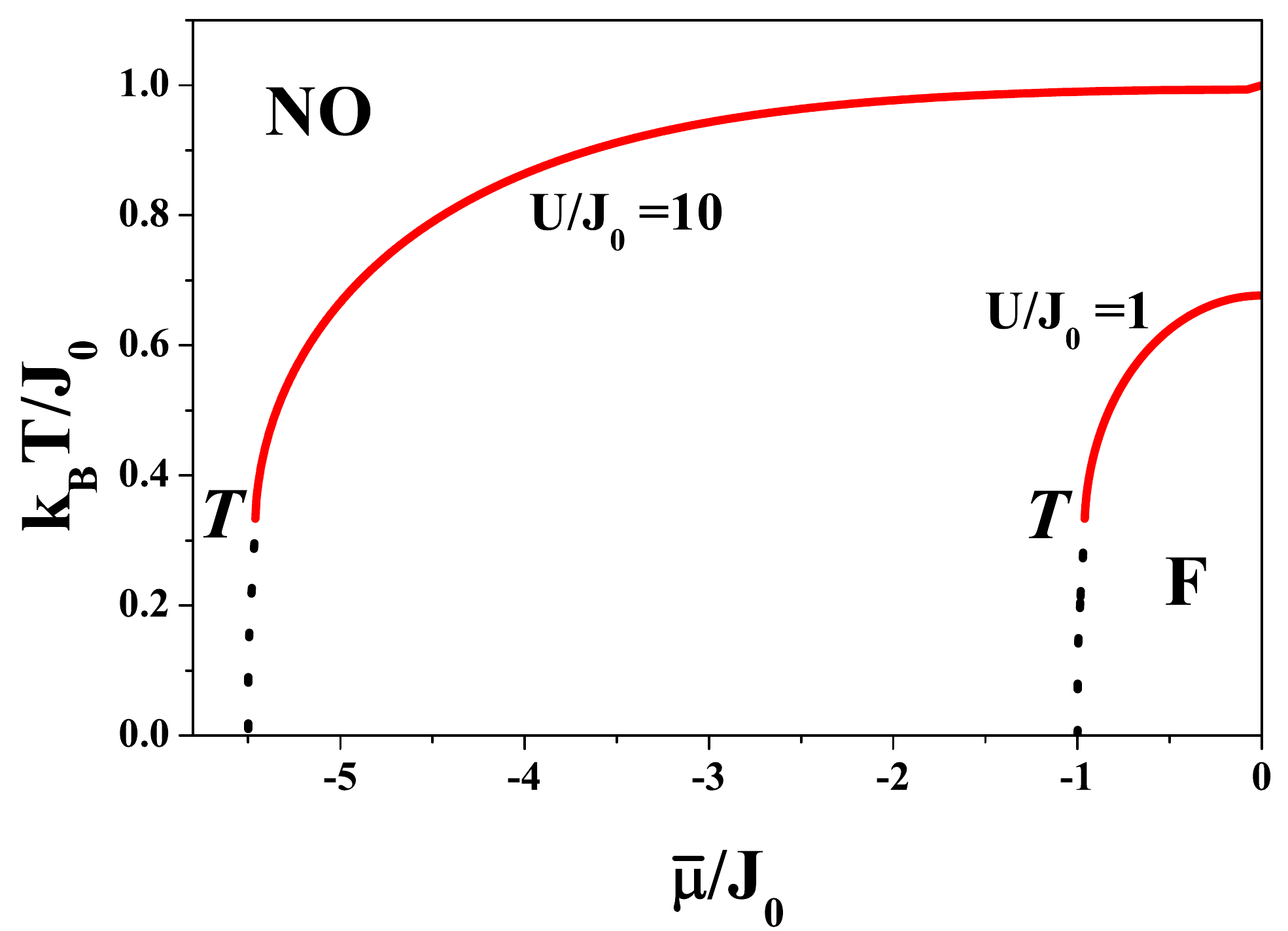}
    \caption{Phase diagrams $k_BT/J_0$ vs. $\bar{\mu}/J_0$ for \mbox{$U/J_0=1,\ 10$} (as labeled) obtained within VA. Dotted and solid lines indicate first and second order boundaries, respectively. $T$ denotes a~tricritical point.}
    \label{rys:fig1}
\end{figure}

Obtained phase diagrams are symmetric with respect to half-filling (\mbox{$n=1$}, \mbox{$\bar{\mu}=\mu-U/2=0$}) because of the particle-hole symmetry of the Hamiltonian (\ref{row:1}), so the diagrams will be presented only in the range \mbox{$0\leq n\leq 1$}.

\section{Results and discussion}

\subsection{The variational approach}

In this subsection we discuss the results for strong on-site repulsion obtained within VA.
The dependencies of the transition temperature \mbox{F--NO} as a function of $\bar{\mu}$ for \mbox{$U/J_0=1$} and  \mbox{$U/J_0=10$} (this is close to the limit \mbox{$U\rightarrow+\infty$}) are shown in Fig.\ref{rys:fig1}.
The range of the F phase stability is reduced with decreasing $U/J_0$. A tricritical point $T$ connected with a~change of the \mbox{F--NO} transition order is located at \mbox{$k_BT/J_0=1/3$} and its location does not dependent on $U/J_0$ in the limit considered.

\begin{figure}
    \centering
    \includegraphics[width=\rozmiar\textwidth]{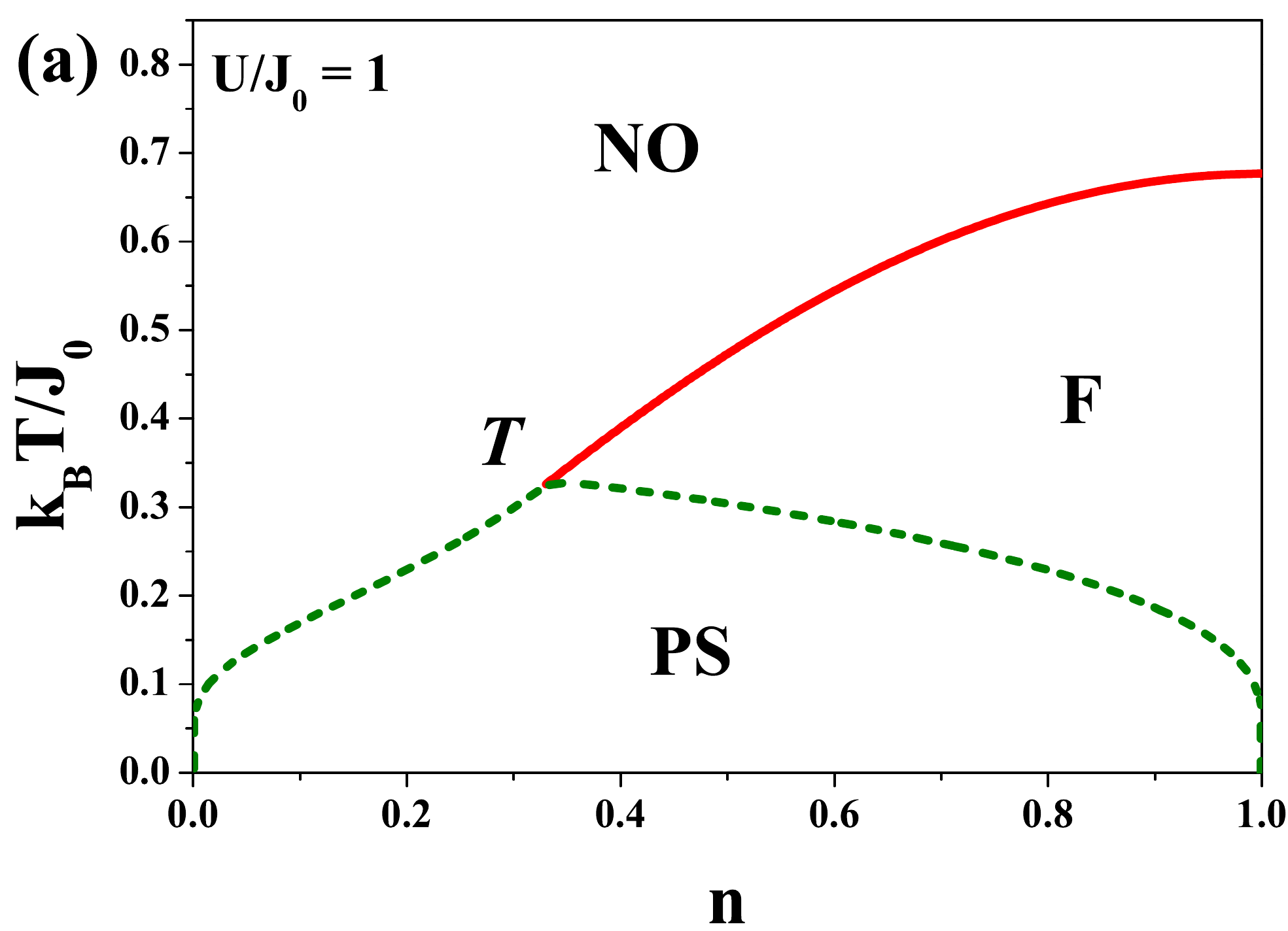}\\
    \includegraphics[width=\rozmiar\textwidth]{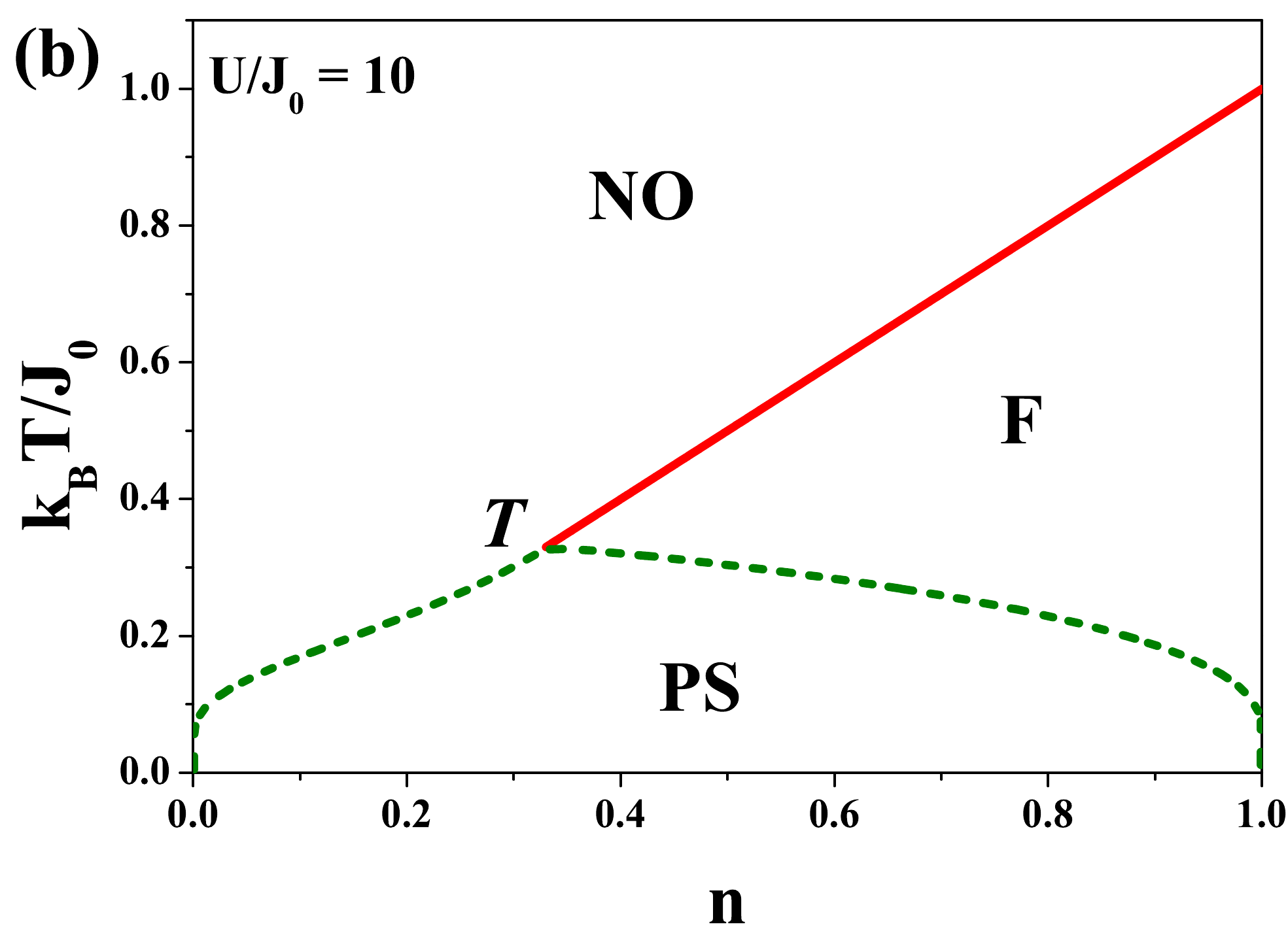}
    \caption{Phase diagrams $k_BT/J_0$~vs.~$n$ for: (a)~\mbox{$U/J_0=1$} and (b)~\mbox{$U/J_0=10$} obtained within VA. Solid and dashed lines indicate second order and ``third order'' boundaries, respectively. $T$ denotes a~tricritical point.}
    \label{rys:fig2}
\end{figure}

If the system is analyzed for fixed~$n$~\cite{KKR2010},
at sufficiently low temperatures the homogeneous phases are not states with the lowest free energy and the PS state can occur. On the phase diagrams, there is a second order line at high temperatures, separating F and NO phases.  A~``third order'' transition takes place at lower temperatures, leading to a PS of the F and NO phases. The critical point for the phase separation ($T$) lies on the second order line \mbox{F--NO} and it is located at \mbox{$k_BT/J_0=1/3$} and \mbox{$n=1/3$}. Phase diagrams for \mbox{$U/J_0=1$} and \mbox{$U/J_0=10$} are shown in Fig.~\ref{rys:fig2}. With increasing $k_BT/J_0$ the system exhibits either a~sequence of transitions: PS$\rightarrow$F$\rightarrow$NO (for \mbox{$1/3<n<1$}) or a~single transition: PS$\rightarrow$NO (for \mbox{$n<1/3$}) and F$\rightarrow$NO (for \mbox{$n=1$}).

\subsection{Monte Carlo results}

\begin{figure}
    \centering
    \includegraphics[width=\rozmiar\textwidth]{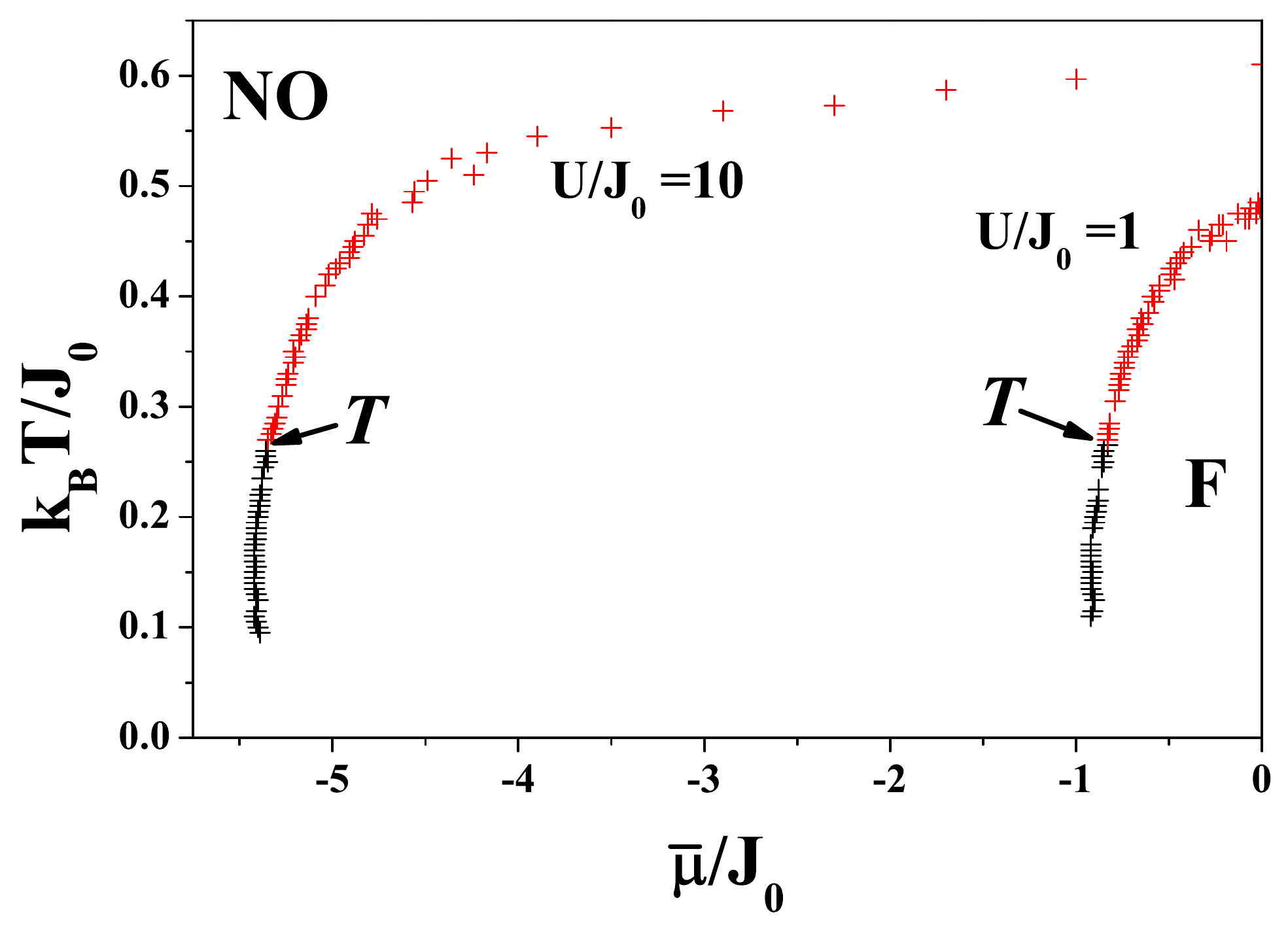}
    \caption{Phase diagrams $k_BT/J_0$ vs. $\bar{\mu}/J_0$ for \mbox{$U/J_0=1,\ 10$} (as labeled) obtained within MC simulation for \mbox{$10 \times 10$} SQ lattice. $T$ denotes a~tricritical point. Details in text.}
    \label{rys:fig3}
\end{figure}

Here, we present a numerical investigations of  model (\ref{row:1}), using standard MC methods in the grand canonical ensemble (for details see e.g. Ref. \onlinecite{P2006}).
The MC simulations have been done for two dimensional SQ lattice (\mbox{$z=4$}) with periodic boundary conditions. The size of the lattice is relatively small, i.e. \mbox{$10 \times 10$}.

The general properties of MC phase diagrams are similar to those obtained within VA. However, it is obvious that the transition temperatures resulting from MC simulations for \mbox{$d=2$} SQ lattice are lower than those obtained within VA, which is exact in the limit of infinite dimensions.

The phase diagrams as a function of $\bar{\mu}$ for \mbox{$U/J_0=1$} and  \mbox{$U/J_0=10$} are shown in Fig.~\ref{rys:fig3}.
The transitions in finite systems are not sharp (the finite-size effect on the order parameter, i.e.~in the NO phase \mbox{$m\neq0$} is larger than zero near the transition) and thus the precise location of boundaries between different phases are determined by the discontinuity of the magnetic susceptibility.  A~tricritical point $T$ connected with a~change of the \mbox{F--NO} transition order is located at \mbox{$k_BT/J_0\approx0.27$}. The \mbox{F--NO} transition can be first order (for temperatures below \mbox{$T$-point}) as well as second order (for temperatures above \mbox{$T$-point}). The maximum for the \mbox{F--NO} transition temperature is at half-filling (\mbox{$n=1$}, \mbox{$\bar{\mu}/J_0=0$}) and it equals (i)~\mbox{$k_BT/J_0\approx0.48$} for \mbox{$U/J_0=1$} and (ii)~\mbox{$k_BT/J_0\approx0.59$} for \mbox{$U/J_0=10$}. The behaviors of the boundaries at low temperatures (i.e. for \mbox{$k_BT/J_0<0.1$}) have not been determined.

\begin{figure}
    \centering
    \includegraphics[width=\rozmiar\textwidth]{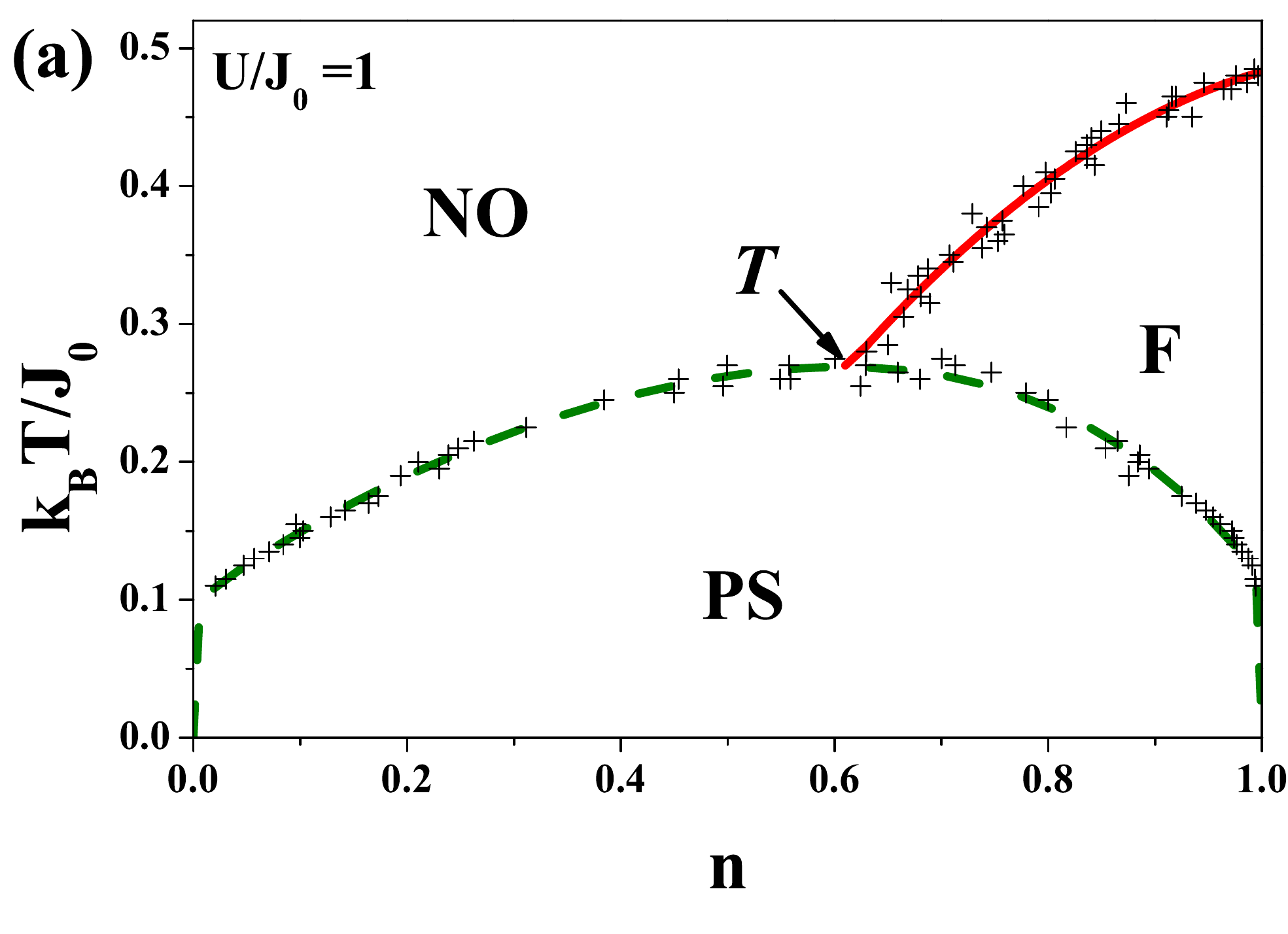}\\
    \includegraphics[width=\rozmiar\textwidth]{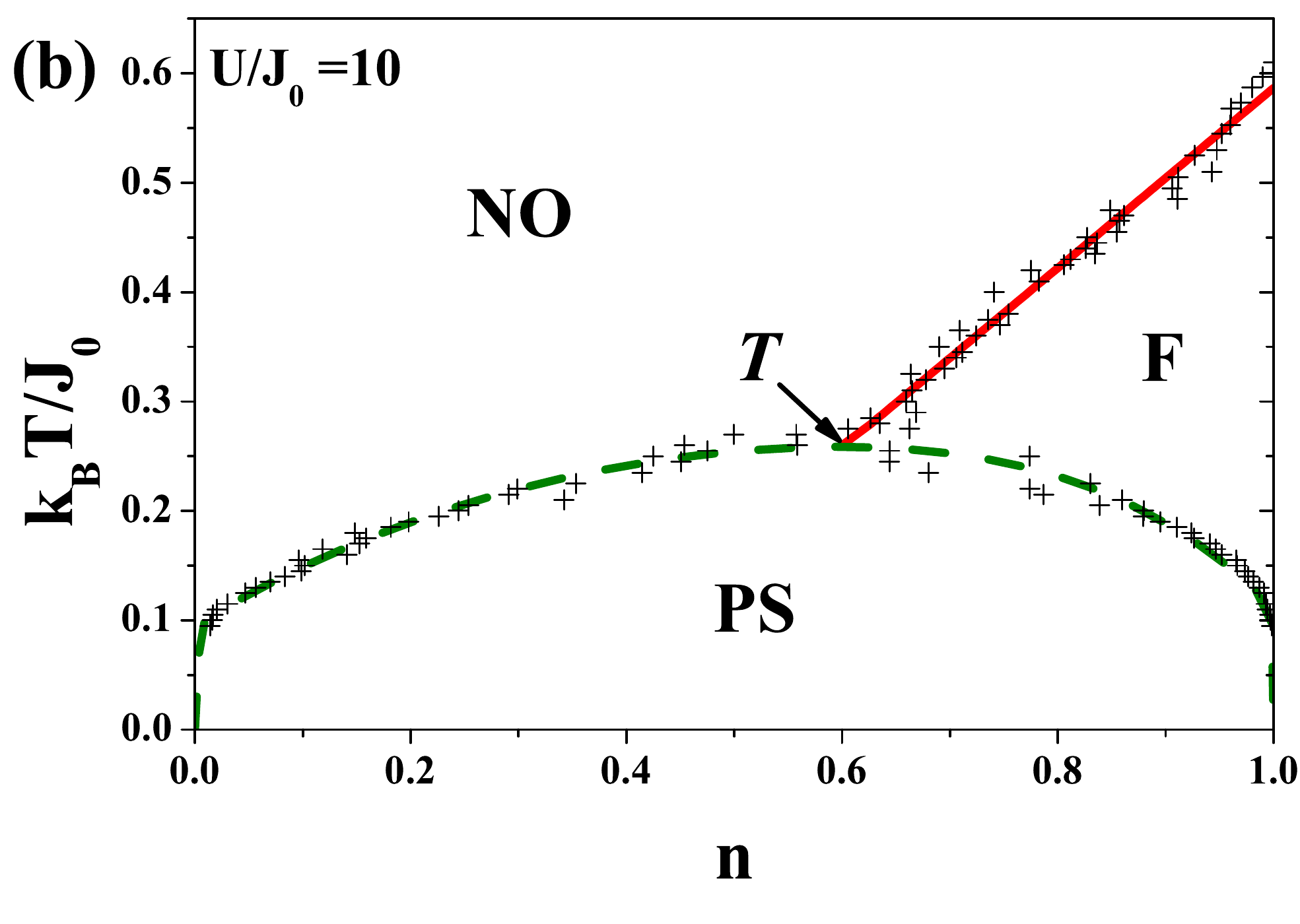}
    \caption{Phase diagrams $k_BT/J_0$~vs.~$n$ for: (a)~\mbox{$U/J_0=1$}  and (b)~\mbox{$U/J_0=10$}  resulting from MC simulation for \mbox{$10 \times 10$} SQ lattice. Denotations as in Fig.~\ref{rys:fig2}. Details in text.}
    \label{rys:fig4}
\end{figure}

One can translated the (grand-canonical) diagrams from Fig.~\ref{rys:fig3} into the (canonical) diagrams for arbitrary $n$ by the standard way. The resulting diagrams are shown in Fig.~\ref{rys:fig4}. At higher temperatures the F and NO phases are separated by a~second order line. At lower temperatures (below \mbox{$T$-point}) the PS state occurs, which is separated from homogeneous phases (i.e. F and NO phases) by ``third order'' boundaries. The tricritical point is placed at substantially  higher electron concentrations (\mbox{$n\approx0.61$}) in comparison to VA results  and (as in VA) its location is independent of the on-site repulsion $U/J_0$ in the limit considered.

\section{Final comments}

We considered a~simple model of magnetically ordered insulators. We presented phase diagrams for strong on-site repulsion including a~tricritical behavior obtained  by Monte Carlo simulations and compared them with VA results. It was shown that MC results are qualitatively similar to those derived within the VA. However, one should notice that the MC transition temperatures are significantly smaller than VA ones.  The \mbox{F--NO} transition can be second as well as first order. At sufficiently low temperatures, where the \mbox{F--NO} transition is discontinuous (if $\bar{\mu}/J_0$ is fixed), homogeneous phases do not exist (if $n$ is fixed) and the phase separated states have a lowest energy.

Let us stress that the knowledge of the zero-bandwidth limit can be used as starting point for a perturbation expansion in powers of the hopping and as an important test for various approximate approaches analyzing the corresponding finite bandwidth models.

We leave the problem of detailed analysis concerning arbitrary $U/J_0$ for the future investigations \cite{KKM0000}.

\begin{acknowledgments}
S.~M. and K.~K. would like to thank the European Commission and Ministry of Science and Higher Education (Poland) for the partial financial support from European Social Fund -- Operational Programme ``Human Capital'' -- POKL.04.01.01-00-133/09-00 -- ``\textit{Proinnowacyjne kszta\l{}cenie, kompetentna kadra, absolwenci przysz\l{}o\'sci}''.
\end{acknowledgments}

\end{document}